# Efficient Denoising Method to Improve The Resolution of Satellite Images

Jhanavi Hegde
*Mountain View High School*
Mountain View, CA, USA

## Abstract

Satellites are widely used to estimate and monitor ground cover, providing critical information to address the challenges posed by climate change. High-resolution satellite images help to identify smaller features on the ground and classification of ground cover types. Small satellites have become very popular recently due to their cost-effectiveness. However, smaller satellites have weaker spatial resolution, and preprocessing using recent generative models made it possible to enhance the resolution of these satellite images. The objective of this paper is to propose computationally efficient guided or image-conditioned denoising diffusion models (DDMs) to perform super-resolution on low-quality images. Denoising based on stochastic ordinary differential equations (ODEs) typically takes hundreds of iterations and it can be reduced using deterministic ODEs. I propose Consistency Models (CM) that utilize deterministic ODEs for efficient denoising and perform super resolution on satellite images. The DOTA v2.0 image dataset that is used to develop object detectors needed for urban planning and ground cover estimation, is used in this project. The Stable Diffusion model is used as the base model, and the DDM in Stable Diffusion is converted into a Consistency Model (CM) using Teacher-Student Distillation to apply deterministic denoising. Stable diffusion with modified CM has successfully improved the resolution of satellite images by a factor of 16, and the computational time was reduced by a factor of 20 compared to stochastic denoising methods. The FID score of low-resolution images improved from 10.0 to 1.9 after increasing the image resolution using my algorithm for consistency models.

## Introduction

Satellites play a crucial role in tracking and understanding climate change by providing images that can be used to monitor ground cover, offering critical information for environmental management, agriculture, forestry, and urban planning. High-resolution satellites like WorldView, GeoEye, and SPOT provide detailed images that can identify smaller features on the ground. These images are essential for precise mapping of ground cover types, such as distinguishing between different types of crops, forest species, or urban features. Figure **1** shows key Climate Change Indicators that can be monitored using satellite data (S. Zhao, 2023):

**Vegetation Monitoring:** Detect changes in forest cover, deforestation, and desertification using NDVI (Normalized Difference Vegetation Index) and other vegetation indices.

**Ice and Snow Cover**: Monitor changes in glaciers, polar ice caps, and snow cover to assess the impact of global warming.

**Urban Expansion**: Detect urban sprawl and changes in land use, which contribute to heat islands and increased carbon emissions.

**Water Bodies**: Monitor changes in the size and health of lakes, rivers, and coastal regions, which can indicate rising sea levels and droughts.

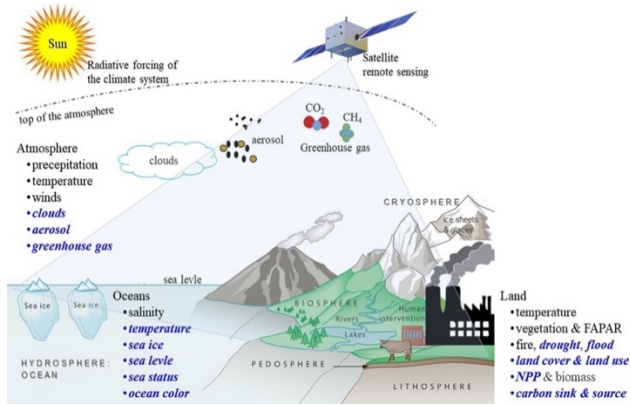

Figure 1 Climate Change Indicators

Computer vision (CV) is increasingly employed to monitor climate change by analyzing satellite data. This approach uses advanced image processing and machine learning techniques to extract critical insights from vast datasets generated by Earth-observing satellites. Machine learning models provide significant advantages over traditional handcrafted CV methods by automatically learning from data, adapting to various tasks, and managing complex patterns and large datasets. Deep learning architectures, in particular, excel in object detection, classification, segmentation, and change detection in satellite images. Given the immense volume of satellite data, machine learning models—especially those designed to handle large datasets—efficiently process and analyze this information at scale, offering substantial time savings with improved performance compared to traditional methods, making them essential for climate monitoring.

There are several practical challenges in using computer vision techniques on satellite images to achieve desired results. The performance of computer vision methods depends on the resolution of images, and deep learning methods, such as Convolutional Neural Networks (CNNs) provide higher accuracy with high-resolution images. Cost-effective smaller satellites are becoming increasingly popular, but they often have lower spatial resolution, limiting the detection of small objects. Additionally, clouds and rain can obstruct the view, making it difficult to capture clear images and challenging to ensure high-quality images across diverse environments and conditions.

Pre-processing is crucial for enhancing the resolution of images to achieve optimal performance from machine learning models after initial image acquisition. Several methods are available to improve the resolution of satellite images, and generative models are increasingly being used for their ability to generate realistic images from degraded ones. DDMs have revolutionized the image Super-Resolution (SR) field, significantly narrowing the gap between image quality and human perceptual preferences (P.Abbeel, 2020). DDMs use conditional image generation and perform super-resolution through a stochastic denoising process. They are relatively easier to train and can produce high-quality samples that surpass the realism of those generated by previous methods such as GANs.

DDMs learn to transform a standard normal distribution into an empirical data distribution through a sequence of refinement steps. DDMs typically use a U-Net model (O. Ronneberger, 2015) that is trained with a denoising objective, iteratively removing varying levels of noise from the output. The SR3 model (C. Saharia, 2021) uses Denoising Diffusion Probabilistic Models (DDPMs) for conditional image denoising by introducing a simple yet effective modification to the U-Net architecture. The SR3 model outperformed traditional regression models and

GANs in producing super-resolution images. However, SR3 requires significant computational resources for denoising in the pixel space and takes several iterations to generate images. Recently, Stable Diffusion (R.Rombach, 2022), a Latent Diffusion Model (LDM) developed by researchers at Ludwig Maximilian University in Munich, implemented the denoising process in a compact latent space, significantly reducing computation. This LDM can be used for various image generation tasks, including super-resolution conditioned on low-resolution images.

Stable Diffusion enables high-quality image synthesis while avoiding excessive computational requirements by training the diffusion model in a compressed, lower-dimensional latent space. Figure 2 shows Stable Diffusion consisting of three components: the Variational Autoencoder (VAE), U-Net, and an optional text or image conditioning module. The VAE encoder compresses the image from pixel space to a smaller-dimensional latent space by removing perceptually irrelevant details and retaining only the fundamental semantic content of the image. This reduction in dimensionality aids in the super-resolution of high-dimensional images. For example, a color image with a 512x512 resolution has 786,432 possible values, while Stable Diffusion compresses it into a latent space that is 16 times smaller, with only 49,152 values. The U-Net model is used for noise removal, and denoising occurs in this compact latent space.

The main steps involved in an LDM (Latent Diffusion Model) are as follows:

**Encoder**: An input image is passed through an encoder model, which compresses it into a smaller latent representation. This latent code captures the most important features and semantics of the image in a compact form.

**Latent Diffusion**: The denoising diffusion process is applied to the latent code rather than directly on the pixels. This allows the DDM to manipulate the image in a more controlled way by only modifying the latent code.

**Decoder**: Once the diffusion process modifies the latent code to generate the desired output image, a decoder model transforms the latent code back into pixel space, reconstructing the final high-resolution image.

LDMs can increase image resolution by 2x, 4x, or even higher by taking a low-resolution image as input, encoding it into the latent space, modifying the latent code, and then decoding it back into pixels. Stable Diffusion uses a Variational Autoencoder (VAE) to encode the image $x_0$ into a smaller-dimensional latent space $z_T$. The encoder preserves key details from the original image, which the decoder then expands into a highly realistic super-resolved version. The U-Net model is responsible for denoising, predicting a denoised image representation from noisy latents $z_T$. The actual latents $z_0$ are obtained by subtracting the predicted noise from the noisy latent $z_T$. This denoising step can be guided flexibly by conditioning it on a text or a blurred variant of the same image. To perform super resolution, the denoising step was conditioned on these low-resolution images. This approach—using a universal autoencoder once and then reusing it for different image generation tasks—is highly effective for improving the quality of satellite images with different resolutions, contrast, and brightness. The pretrained Stable Diffusion model is used as a base and fine-tuned for enhancing satellite image quality.

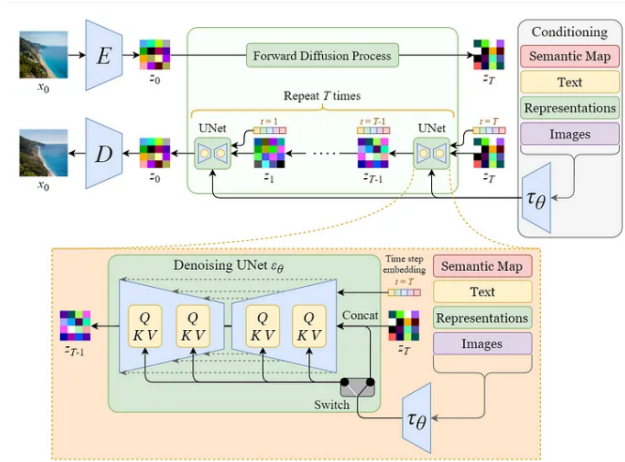

Figure 2. Stable Diffusion

The scope of this project is limited to modifying the DDM in Stable Diffusion to reduce computational time during denoising and implementing an inference pipeline by using the existing decoder to convert latents into RGB images. It is important to explain how the DDM works, followed by a description of the improvements made to the denoising process. The three main aspects of the DDM are outlined below:

**Forward Diffusion Process**: The model first corrupts the data by adding noise over several time steps. This is a Markovian process, where at each time step, Gaussian noise is added to the data, progressively degrading it into pure noise. This process can be described mathematically as:

$$q(x_t|x_{t-1}) = \mathcal{N}(x_t; \sqrt{\alpha_t}x_{t-1}, (1 - \alpha_t)\mathbf{I})$$

Where is $x_t$ is the noisy version of the data at step t and $\alpha_t$ controls the noise level at each step. After many steps, the data becomes indistinguishable from random noise.

**Reverse Diffusion Process**: The goal of the model is to reverse this noising process: given a noisy sample, predict a less noisy version of it, eventually recovering the original data. The reverse process is parameterized by a neural network that learns to denoise the noisy data step by step. The probability distribution in the reverse process is also Markovian and given by:

$$p_\theta(x_{t-1}|x_t) = \mathcal{N}(x_{t-1}; \mu_\theta(x_t, t), \Sigma_\theta(x_t, t))$$

Where $\mu_\theta(x_t, t)$ and $\Sigma_\theta(x_t, t)$ are the mean and variance predicted by the neural network. The forward and reverse diffusion processes are shown in Figure 4 (P.Abbeel, 2020).

**Training Objective:** DDPM is trained by loss function that looks like the mean squared error (MSE) between the added noise and the noise predicted by the model at each diffusion step. This loss at each timestep is:

$$\mathcal{L}_t = \mathbb{E}_{x_0, \epsilon, t} \left[ \|\epsilon - \epsilon_\theta(x_t, t)\|^2 \right]$$

where $x_0$ is the original data ε is the Gaussian noise added in the forward process, and $\varepsilon_\theta(x_t, t)$ is the noise predicted by the neural network.

The training and sampling (i.e., image generation) algorithm of DDPM are given below in Figure 3. Stable Diffusion uses the same method as in SR3 for Image conditioned denoising as shown Figure 5 in (C. Saharia, 2021).

The low-resolution image $x$ is concatenated with noisy image $y_t$ and combined images are used as input to U-Net model. Predicted noise is subtracted from $y_t$ to generate $y_{t-1}$. Successive denoising continue to use low-resolution image $x$ to guide denoising by concatenating with noisy images. In Stable Diffusion, the noisy images $y_t$ are replaced by latents $z_t$. To improve the speed of denoising, we need to simplify the sampling method used for the reverse diffusion process.

**Algorithm 1** Training
1: **repeat**
2: $\quad \mathbf{x}_0 \sim q(\mathbf{x}_0)$
3: $\quad t \sim \text{Uniform}(\{1, \ldots, T\})$
4: $\quad \epsilon \sim \mathcal{N}(\mathbf{0}, \mathbf{I})$
5: $\quad$ Take gradient descent step on
$\quad\quad \nabla_\theta \left\| \epsilon - \epsilon_\theta(\sqrt{\bar{\alpha}_t}\mathbf{x}_0 + \sqrt{1-\bar{\alpha}_t}\epsilon, t) \right\|^2$
6: **until** converged

**Algorithm 2** Sampling
1: $\mathbf{x}_T \sim \mathcal{N}(\mathbf{0}, \mathbf{I})$
2: **for** $t = T, \ldots, 1$ **do**
3: $\quad \mathbf{z} \sim \mathcal{N}(\mathbf{0}, \mathbf{I})$ if $t > 1$, else $\mathbf{z} = \mathbf{0}$
4: $\quad \mathbf{x}_{t-1} = \frac{1}{\sqrt{\alpha_t}} \left( \mathbf{x}_t - \frac{1-\alpha_t}{\sqrt{1-\bar{\alpha}_t}} \epsilon_\theta(\mathbf{x}_t, t) \right) + \sigma_t \mathbf{z}$
5: **end for**
6: **return** $\mathbf{x}_0$

Figure 3. DDIM Algorithm

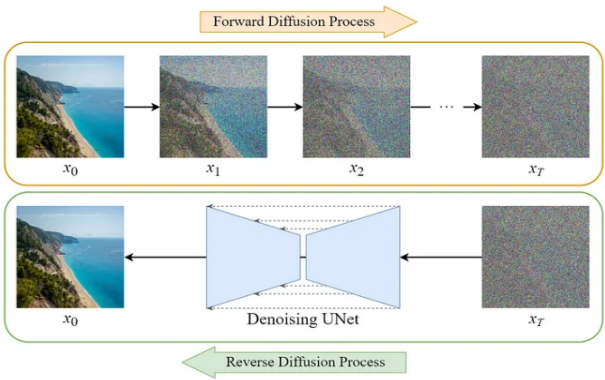

Figure 4. Denoising Diffusion

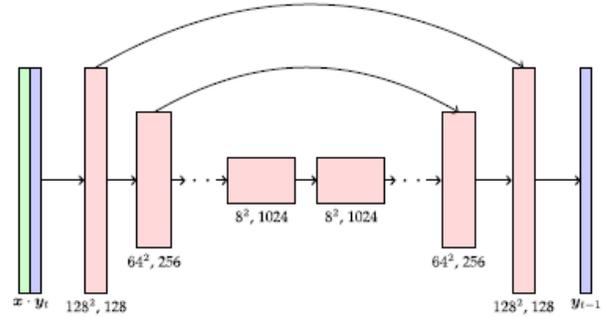

Figure 5. Image Conditioned U-Net

The DDPM used in Stable Diffusion employs the same number of steps in the reverse process as in the forward process. Denoising low-resolution images with the DDPM algorithm is very slow, as it takes hundreds of steps. For real-time applications, reducing time and computational resources is crucial. Denoising Diffusion Implicit Model (DDIM) was proposed to accelerate image generation (J. Song, 2022) and the denoising process in DDIM is shown in Figure 6, where the number of steps is reduced by sampling every nth step (with n > 1). In the example below, the value of n is 2.

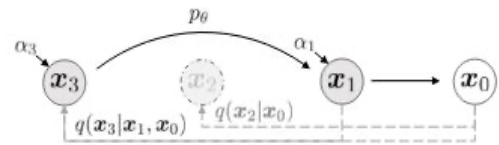

Graphical model for accelerated generation, where $\tau = [1, 3]$.

Figure 6. Denoisng diffusion implicit models

However, image quality deteriorates for large values of n, as the denoising trajectories remain stochastic. Recently proposed Consistency Models (CMs), directly map low-resolution images to high-resolution data and allow the use of deterministic denoising methods, as shown in Figure 7 (Y. Song, 2023).

The CM paper (Y. Song, 2023) proposed two methods for training Consistency Models, and in this project, Consistency Distillation (CD) method is used for training, where the CM is trained as a student model from a pre-trained teacher model. The training objective is to minimize the difference between the model outputs for pairs generated from the same trajectory.

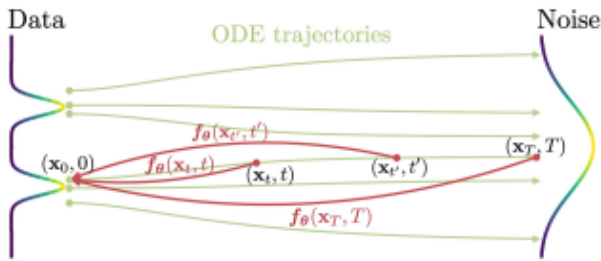

Figure 7. Deterministic Trajectory

The scope of this project is to train a DDM to use computationally efficient denoising based on deterministic trajectory and generate high-resolution images from low-resolution satellite images. The method of training CM model and implementing the inference pipeline for satellite images will be discussed in the next section.

DOTA images are used as training and test datasets. Even though the size of the training and test datasets are quite small compared to typical size of datasets used for training generative models, the size is large enough to demonstrate improvement in denoising time using CMs.

Google Colab platform with V100 GPU was used for training the CMs and running inference pipeline to generate results. The source code in Huggingface diffusers repository is used as a baseline for this project. The CM training for text to image generation example in the repository is modified to support guided image generation conditioned on low resolution images. A new inference pipeline was implemented to perform super resolution using CM.

The PyTorch implementation of CM training and the inference pipeline are available at - *https://github.com/jhanavi-h/diffusers/tree/dev_sr_branch/*.

**Methods**

This method offers an advantage when a pre-trained model is already available from Hugging Face (HuggingFace, 2022), making the implementation more efficient. Algorithm of Teacher–Student Distillation process is described in Figure 8 (Y. Song, 2023).

**Algorithm 2** Consistency Distillation (CD)
**Input:** dataset $\mathcal{D}$, initial model parameter $\theta$, learning rate $\eta$, ODE solver $\Phi(\cdot,\cdot;\phi)$, $d(\cdot,\cdot)$, $\lambda(\cdot)$, and $\mu$
$\theta^- \leftarrow \theta$
**repeat**
  Sample $\mathbf{x} \sim \mathcal{D}$ and $n \sim \mathcal{U}[\![1, N-1]\!]$
  Sample $\mathbf{x}_{t_{n+1}} \sim \mathcal{N}(\mathbf{x}; t_{n+1}^2 \mathbf{I})$
  $\hat{\mathbf{x}}_{t_n}^\phi \leftarrow \mathbf{x}_{t_{n+1}} + (t_n - t_{n+1})\Phi(\mathbf{x}_{t_{n+1}}, t_{n+1}; \phi)$
  $\mathcal{L}(\theta, \theta^-; \phi) \leftarrow$
    $\lambda(t_n) d(f_\theta(\mathbf{x}_{t_{n+1}}, t_{n+1}), f_{\theta^-}(\hat{\mathbf{x}}_{t_n}^\phi, t_n))$
  $\theta \leftarrow \theta - \eta \nabla_\theta \mathcal{L}(\theta, \theta^-; \phi)$
  $\theta^- \leftarrow \text{stopgrad}(\mu \theta^- + (1-\mu)\theta)$
**until** convergence

Figure 8. Consistency Distillation Training

The CM $f_\theta(x,t)$ is characterized by the equation given below where it is a unity function at $t = \epsilon$.

$$f_\theta(\mathbf{x}, t) = \begin{cases} \mathbf{x} & t = \epsilon \\ F_\theta(\mathbf{x}, t) & t \in (\epsilon, T] \end{cases}.$$

The pre-trained neural network $F_\theta(x,t)$ provides the trajectory for training the CM $f_\theta(x,t)$. The CM is trained over the time horizon

($\epsilon, T$) with $N < T$ sub samples, at time instances $t_1 = \epsilon < t_2 .. < t_{N-1} < t_N = T$. Given a datapoint $x$, the noisy sample $x_{t_{n+1}}$ is generated by adding noise with variance proportional to $t_{n+1}^2$ to $x$. Then a pair of adjacent data points $\left(x_{t_n}^\emptyset, x_{t_{n+1}}\right)$ are generated where $x_{t_n}^\emptyset$ is generated by one step discretization step from the output $x_{t_{n+1}}$ from the teacher model. Then the student model $f_\theta(x, t)$ is trained using back propagation to minimize the mean square distance - $d\left(f_\theta\left(x_{t_n}^\emptyset\right), f_\theta(x_{t_{n+1}})\right)$. The pretrained DDIM from Huggingface is used as teacher model and CM training is done with mini-batch size of 12 with $n$ values sampled from $U[1, N-1]$. The CD training method in the CM paper (Y. Song, 2023) is modified to use guided image generation by conditioning the latents on low-resolution satellite images, as done in the SR3 implementation (C. Saharia, 2021). The CM paper (Y. Song, 2023) proposed sampling or image generation using a greedy algorithm, but this project proposed much simpler DDIM sampling that conditions the sampling process on the original image to remove Markovian dependency. This allows sampling with larger step sizes and enables super-resolution in just 3 to 4 steps.

**Dataset**

The DOTA v2.0 image dataset (Daa, 2018) is used for this project. DOTA consists of images collected from the Google Earth, GF-2, JL-1 satellites and aerial imagery provided by CycloMedia B.V. DOTA images are used to develop and evaluate object detectors needed for urban planning and ground cover estimation. The training data consists of 2,800 images, while the test data contains 200 images from the DOTA dataset. Even though the size of the training and test datasets are quite small compared to typical size of datasets used for training generative models, the size is large enough to demonstrate improvement in denoising time using CMs. The dataset has both RGB and grayscale images of varying sizes, all stored in PNG format. The images are labeled with various objects, but this information is ignored for the purposes of super-resolution. The ETL (extract, transform, load) pipeline selects only the RGB images, cropping them to a uniform size of 512 x 512 pixel. These cropped images serve as the reference for training and performance evaluation. Additionally, the 512 x 512 images are down sampled by a factor of 4 to create low-resolution counterparts of size 128 x 128 pixels. The LDM is expected to enhance the resolution by a factor of 16, with performance assessed by comparing the model's output to the original 512 x 512 image.

**Results**

The consistency distillation example in the Huggingface Diffusers repository was modified to support guided image generation conditioned on low-resolution images. These changes were necessary to convert text-to-image generation into image-to-image generation. The training parameters for CD training are provided in Table 1.

| Hyper-Parameter | Value |
|---|---|
| Learning Rate | 0.000001 |
| Loss Type | Huber |
| N | 50 |
| T | 1000 |
| Batch Size | 12 |
| Image Size | 512x512 |
| Time Step Scaling Factor | 10 |

Table 1 Hyper-parameters

The hyperparameters of CD training were not changed, except for the batch size, which was adjusted to fit the training dataset into GPU memory. Stable Diffusion uses $T = 1000$ steps for forward diffusion process and the variance of the added noise is varied from $\beta_1 = 10^{-4}$ to from $\beta_T = 0.02$ and image data is normalized to $[-1, 1]$. The number of sub-samples, $N = 50$ is used for CD training. The batch size is set to 12 and limited by the GPU RAM size. Figure **9** shows plot of Huber loss of CD training, and the learning curve does not show graceful reduction with number of samples. The training loss is small even at the beginning because the student model is initialized with same weights as pretrained teacher model.

It was difficult to judge the convergence of the model based on the training loss and the training loss was stopped after one epoch. There was no improvement seen by increasing the number of epochs. The stopping criteria for Teacher Student Distillation is not clear at this time and to be studied in the future.

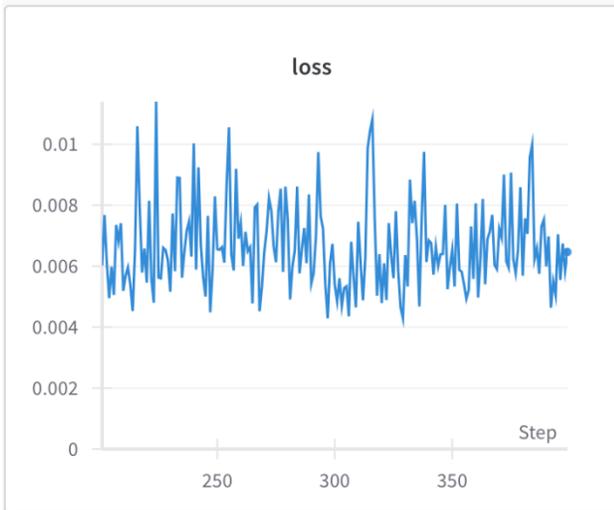

Figure 9 Training Loss

The Super Resolution process using Consistency Models (CM) is expected to be faster than DDPM, reducing the number of denoising steps from $T = 1000$ to 4, while maintaining comparable image quality. Traditional metrics like Mean Square Error (MSE) and Peak Signal-to-Noise Ratio (PSNR) are not ideal for evaluating generative models. Although high-resolution images produced by CM show only a modest improvement in PSNR (up to around 4 dB), they exhibit substantial enhancements in Frechet Inception Distance (FID) scores. FID measures the similarity between the distribution of generated images and real images, with lower scores indicating more realistic outputs. Unlike MSE, which compares images pixel by pixel, FID evaluates the mean and standard deviation of activations in the deepest layers of the Inception v3 model. The FID score improved from 10 to 1.9 after improving the resolution using the proposed CM method. Lower FID scores imply improved feature extraction when enhanced satellite images are used in CNN-based object detection models for monitoring climate change. PSNR was improved by more than 3 dB for over 60% of the images.

| Metric | Performance |
|---|---|
| PSNR | > 3 dB for 60% images |
| FID | 1.9 (With Super Resolution) |
|  | 10.0 (Before at Low Resolution) |

Table 2 Improvement in Image Quality

The perceptual quality was significantly enhanced, as seen in the two examples given below.

Figure 10 (Right: Original, Left: Improved)

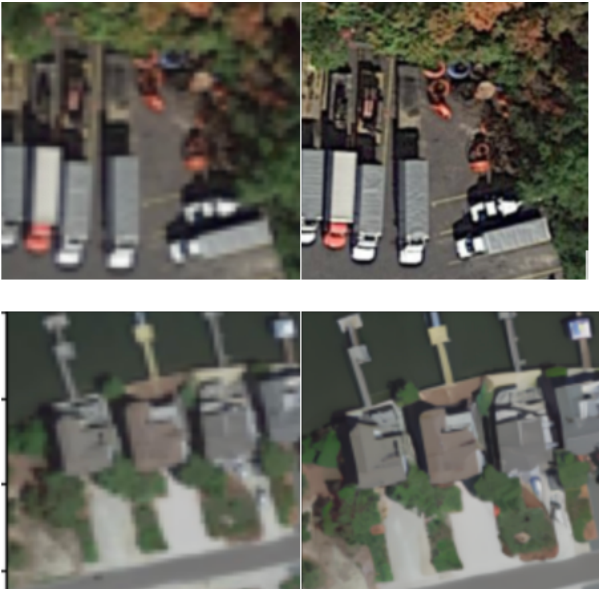

More examples of images with improved resolutions are available at my GitHub repo - *https://github.com/jhanavi-h/diffusers/tree/main/images*.

## Conclusion

The results indicate that modifying Stable Diffusion to use CM as a deterministic denoising model can significantly reduce computational resources and time. This model is a strong candidate for performing super-resolution on satellite images. Improved FID scores for super-resolution demonstrate that this implementation can be highly beneficial in the preprocessing module of satellites, enhancing the performance of downstream tasks such as object detection and classification, which are essential for tracking climate change.

The deterministic denoising method of CM demonstrates significant improvement in the perceptual quality of DOTA images, and it is worthwhile to apply this method and evaluate its performance on larger commercial datasets (e.g., Sentinel-2). However, further improvements in the training algorithm are needed to generalize across larger datasets. The stopping criteria for improving the convergence of Teacher-Student Distillation training remains an active area of research, and adjustments will be necessary as better training methods become available. I am collaborating with the Huggingface AI community to release my modifications to the open-source community.